\documentstyle[12pt]{article}  
\def\sq{\hbox {\rlap{$\sqcap$}$\sqcup$}}
\def\sq{\hbox {\rlap{$\sqcap$}$\sqcup$}}
\def\R{ {\rm R \kern -.31cm I \kern .15cm}}
\def\C{ {\rm C \kern -.15cm \vrule width.5pt \kern .12cm}}
\def\Z{ {\rm Z \kern -.27cm \angle \kern .02cm}}
\def\N{ {\rm N \kern -.26cm \vrule width.4pt \kern .10cm}}
\def\1{{\rm 1\mskip-4.5mu l} }
\def\lsim{\raise0.3ex\hbox{$<$\kern-0.75em\raise-1.1ex\hbox{$\sim$}}}
\def\gsim{\raise0.3ex\hbox{$>$\kern-0.75em\raise-1.1ex\hbox{$\sim$}}}
\def\noi{\noindent}

\def\beq{\begin{equation}}   \def\eeq{\end{equation}}
\def\bea{\begin{eqnarray}}  \def\eea{\end{eqnarray}}
\def\nn{\nonumber}
\def\noi{\noindent}
\def\beeq{\begin{eqnarray}} \def\eeeq{\end{eqnarray}}
\newcommand\mysection{\setcounter{equation}{0}\section}

\newcounter{hran}

\begin{document} 
\centerline{\large\bf Partially classical limit of the Nelson model} 
 \vskip 3 truemm \centerline{\large\bf } 

\vskip 0.5 truecm

\centerline{\bf J. Ginibre}
\centerline{Laboratoire de Physique Th\'eorique\footnote{Unit\'e Mixte de
Recherche (CNRS) UMR 8627}}  \centerline{Universit\'e de Paris XI, B\^atiment
210, F-91405 ORSAY Cedex, France}
\vskip 3 truemm
\centerline{\bf F. Nironi, G. Velo}
\centerline{Dipartimento di Fisica, Universit\`a di Bologna}  \centerline{and INFN, Sezione di
Bologna, Italy}

\vskip 1 truecm

\begin{abstract}
We consider the Nelson model which describes a quantum system of
nonrelativistic identical particles coupled to a possibly massless
scalar Bose field through a Yukawa type interaction. We study the
limiting behaviour of that model in a situation where the number of
Bose excitations becomes infinite while the coupling constant tends to
zero in a suitable sense. In that limit the appropriately rescaled Bose
field converges to a classical solution of the free wave or
Klein-Gordon equation depending on whether the mass of the field is
zero or not, the quantum fluctuations around that solution satisfy the
wave or Klein-Gordon equation and the evolution of the
nonrelativistic particles is governed by a quantum dynamics with an
external potential given by the previous classical solution.
\end{abstract}

\vskip 3 truecm
\noi AMS Classification : Primary 81T27. Secondary 81T10.  \par \vskip 2 truemm

\noi Key words : Nelson model, classical limit.\par 
\vskip 1 truecm

\noindent LPT Orsay 04-116\par
\noindent November 2004\par

\newpage
\pagestyle{plain}
\baselineskip 18pt

\mysection{Introduction}
\hspace*{\parindent} Quantum theories are generally expected to reduce to the corresponding
classical ones when suitable parameters converge to a limit which is
usually taken to be zero. In ordinary quantum mechanics this parameter
is identified with Planck's constant $\hbar$. The comparison between
those two types of theories was first considered by Schr\"odinger
\cite{9r} and by Ehrenfest \cite{1r} for simple systems with a finite
number of degrees of freedom, and later put on a firm mathematical
basis by Hepp \cite{6r} for more general systems including some with an
infinite number of degrees of freedom. The study of the transition from
quantum descriptions to classical descriptions is a quite active field
of research. However, while the literature concerning systems with a
finite number of degrees of freedom is rather extensive, in the case of
an infinite number of degrees of freedom very few examples have been
analyzed in reasonable depth \cite{6r} \cite{4r} \cite{3r}.\par

In this paper we consider that problem for the so called Nelson model,
which describes a quantum system of nonrelativistic identical particles
interacting with a real scalar field in space-time ${I\hskip-1truemm
R}^{3+1}$. In the formalism of second quantization for the particles
the Hamiltonian of the system is taken to be

\beq \label{1.1e}
H(\psi , a) = (2M)^{-1} \int dx (\nabla \psi )^* (\nabla \psi ) + \int dk\ \omega\ a^*a + \lambda \int dx\ \varphi\ \psi^*\psi
\eeq

\noi where $\omega (k) = (k^2 + \mu^2)^{1/2}$ with $\mu \geq 0$ ($\mu$
is the mass of the bosons), $\psi$, $\psi^*$, $a$, $a^*$ are Heisenberg
field operators satisfying 
\beq \label{1.2e} \left \{ \begin{array}{l} \left [ \psi (t, x) , \psi
(t, x') \right ]_{\mp} = 0\\ \\ \left [ \psi (t, x) , \psi^* (t, x')
\right ]_{\mp} = \delta (x-x') \end{array} \right .\eeq
\beq \label{1.3e} \left \{ \begin{array}{l} \left [ a (t, k) , a
(t, k') \right ]_{-} = 0\\ \\ \left [ a (t, k) , a^* (t, k')
\right ]_{-} = \delta (k-k') \end{array} \right .\eeq

\noi and
\beq
\label{1.4e}
\varphi (t, x) = (2 \pi )^{-3/2} \int dk \ (2 \omega (k))^{-1/2} \left ( a(t, k) e^{ik\cdot x} + a^*(t, k) e^{-ik\cdot x}\right ) \ .
\eeq

\noi The $-$ sign in (\ref{1.2e}) and (\ref{1.3e}) denotes commutators
and the $+$ sign in (\ref{1.2e}) denotes anticommutators. The field
$\psi$ can be either a boson or a fermion field. The time evolution of
the fields $\psi$ and $a$ in the Heisenberg picture is given by the
equations of motion
\beq \label{1.5e} \left \{ \begin{array}{l} i \partial_t \psi = \left [ \psi , H
\right ]_{-} \\ \\ i\partial_t a = \left [ a, H
\right ]_{-}\end{array} \right . 
 \eeq

\noi which, using (\ref{1.1e}), can be explicitly written as
\beq
\label{1.6e}
i \partial_t \psi = - (2M)^{-1} \Delta \psi + \lambda \varphi \psi
\eeq
\beq
\label{1.7e}
i \partial_t a = \omega a + \lambda (2 \omega)^{-1/2} F(\psi^*\psi )
\eeq

\noi where $F$ denotes the Fourier transform. The initial conditions
are denoted by\break\noindent $\psi (t= 0) = \psi_0$, $a(t=0) = a_0$ and $\varphi (t=0)
= \varphi_0$. In the same vein as in \cite{4r} we want to study the
classical limit of the scalar field, keeping however intact the quantum
nature of the nonrelativistic particles. In the following we present a
heuristic discussion of the problem which underlies the rigorous
developments of the next sections.\par

The classical limit is obtained by considering the average of the field
operators on a sequence of states which contain a number $n$ of scalar
particles increasing to infinity. The traditional way to construct such
a sequence is through the Weyl operators
\beq
\label{1.8e}
C(\alpha ) = \exp \left ( \int dk \left (a_0^* \alpha - a_0 \overline{\alpha}\right ) \right )
\eeq

\noi which applied to the Fock vacuum of the scalar particles generate
the coherent states for the operators $(a_0, a_0^*)$. The sequence of
operators $C(n^{1/2}\alpha )$, where $n$ is a positive integer, applied
to any fixed state meets the requirements of the previously mentioned
sequence of states. The average of $\varphi$ on such states scales as
$n^{1/2}$ so that, in order to obtain a finite non trivial limiting
equation for (\ref{1.6e}) when $n$ converges to infinity, we need to
relate $\lambda$ to $n$ according to $n = \lambda^{-2}$. Therefore this
classical limit is at the same time a weak coupling limit. From now on
we shall use $\lambda$ as a parameter which will eventually tend to
zero. We introduce a real function of space-time $A$ conveniently
written in Fourier transform as
\beq
\label{1.9e}
A(t,x) = (2\pi)^{-3/2} \int dk \ (2 \omega (k))^{-1/2} \left ( \alpha (t, k) e^{ik\cdot x} + \overline{\alpha (t, k)} \ e^{-ik\cdot x}\right ) \ ,
\eeq

\noi to be thought as the limit of the rescaled field $\lambda \varphi$
when $\lambda$ tends to zero. The equation of motion (\ref{1.7e}) for
$a$ can be trivially rewritten as
\beq
\label{1.10e}
i \partial_t \alpha + i \partial_t (\lambda a- \alpha ) = \omega \alpha + \omega (\lambda a - \alpha ) + \lambda^2(2 \omega )^{-1/2} F(\psi^*\psi ) \ .
\eeq 

\noi In order to obtain a non trivial limit for (\ref{1.10e}) when
$\lambda$ converges to zero, we impose on $\alpha$ to be solution of
the equation 
\beq
\label{1.11e}
i\partial_t \alpha = \omega \alpha \ .
\eeq 

\noi By rewriting the equations (\ref{1.5e}) in terms of the variables
$\psi$ and $a - \alpha_{\lambda}$, with $\alpha_{\lambda}(t, k) =
\lambda^{-1} \alpha (t, k)$, we obtain 
\beq
\label{1.12e}
i \partial_t \psi = [\psi, K]_-
\eeq
\beq
\label{1.13e}
i \partial_t \left ( a - \alpha_{\lambda}\right ) = \left [ a - \alpha_{\lambda}, K \right ]_-
\eeq

\noi where
\bea
\label{1.14e}
&&K = (2M)^{-1} \int dx \ (\nabla \psi)^* (\nabla \psi) + \int dk\ \omega \left ( a - \alpha_{\lambda}\right )^* \left ( a - \alpha_{\lambda} \right ) + \nn \\
&&\qquad + \int dx\ A \ \psi^*\psi + \int dx (\lambda \varphi - A) \psi^*\psi \ .
\eea

\noi In order to study the limit of (\ref{1.12e}) and (\ref{1.13e})
when $\lambda$ tends to zero we have to change the initial conditions
for $a - \alpha_{\lambda}$. We define the new field variables $\theta
(t)$ and $b(t)$ by
\beq
\label{1.15e}
\theta (t) = C\left ( \alpha_{\lambda}(0)\right )^* \psi (t) \ C \left ( \alpha_{\lambda} (0) \right )
\eeq 
\beq
\label{1.16e}
b (t) = C\left ( \alpha_{\lambda}(0)\right )^* \left ( a(t) -  \alpha_{\lambda}(t) \right ) C \left ( \alpha_{\lambda} (0) \right )
\eeq 

\noi so that
\bea
\label{1.17e}
&&\theta (0) = \psi_0\nn \\
&&b(0) = a_0 \ .
\eea

\noi The $b$'s are the quantum fluctuations around the classical
solution $\alpha$. The equations (\ref{1.12e}) and (\ref{1.13e}) take
the form
\bea
\label{1.18e}
&&i \partial_t \theta = [\theta , L]_-\nn \\
&&i \partial_t b = [b , L]_-
\eea

\noi where 
\bea
\label{1.19e}
&&L = (2M)^{-1} \int dx (\nabla \theta)^* (\nabla \theta ) + \int dk\ \omega \ b^*b + \int dx\ A \ \theta^*\theta \nn \\
&&\quad + \lambda \int dx\ \theta^*\theta \left ( F^{-1} (2 \omega)^{-1/2} b + F(2 \omega )^{-1/2} b^*\right ) \ .
\eea

\noi In (\ref{1.19e}) the only term containing explicitly $\lambda$ is
expected to converge to zero with $\lambda$ so that the putative
limiting equations of (\ref{1.18e}) become
\beq
\label{1.20e}
\left \{ \begin{array}{l} i \partial_t \theta ' = - (2M)^{-1} \Delta \theta ' + A \theta '\\ \\ i \partial_t b' = \omega b' \ . \end{array} \right .
\eeq

\noi The system (\ref{1.18e}) subject to the initial condition
(\ref{1.17e}) is conveniently solved by using the transformation which
connects the Schr\"odinger picture to the Heisenberg picture. The
solution can be written as
\beq
\label{1.21e}
\left \{ \begin{array}{l} \theta (t) = W(t)^* \ \psi_0 \ W(t) \\ \\ b(t) = W(t)^* \ a_0 \ W(t) \end{array} \right .
\eeq

\noi where $W(t)$ is the unitarity propagator satisfying
\bea
\label{1.22e}
i \partial_t W(t) &=& \left \{ (2M)^{-1} \int dx (\nabla \psi_0)^* (\nabla \psi_0) + \int dk \ \omega\ a_0^* a_0\right . \nn \\
&&\left . + \int dx\ A\ \psi_0^*\psi_0 + \lambda \int dx\ \varphi_0 \psi_0^*\psi_0 \right \} W(t)
\eea

\noi and $W(0) = \1$. The Schr\"odinger propagator $W(t)$ is expressed
in terms of the Schr\"odinger field operators, which coincide with the
Heisenberg field operators at time $t = 0$. Similarly, the solution of
the limiting system (\ref{1.20e}) subject to the conditions $\theta '
(0) = \psi_0$ and $b'(0) = a_0$ is given by
\beq
\label{1.23e}
\left \{ \begin{array}{l} \theta ' (t) = V(t)^* \ \psi_0\ V(t) \\ \\ b'(t) = V(t)^* \ a_0 \ V(t)\end{array} \right .
\eeq

\noi where $V(t)$ is the unitary propagator satisfying 
\beq
\label{1.24e}
i \partial_t \ V(t) = \left \{ (2M)^{-1} \int dx (\nabla \psi_0)^* (\nabla \psi_0) + \int dk\ \omega\ a_0^*a_0 + \int dx \ A\ \psi_0^*\psi_0 \right \} V(t)
\eeq 

\noi and $V(0) = \1$. It can be checked directly that 
\beq
\label{1.25e}
W(t) = C \left ( \alpha_{\lambda}(t)\right )^* \ U(t)\ C \left ( \alpha_{\lambda}(0)\right )
\eeq

\noi where
\beq
\label{1.26e}
U(t) = \exp \left ( - i t H(\psi_0, a_0)\right ) \ .
\eeq

\noi Now the previous perturbation problem in the coupling constant
$\lambda$ reduces to comparing the two families of operators $W(t)$ and
$V(t)$ and to showing that
\beq
\label{1.27e}
\lim_{\lambda \to 0} W(t) = V(t) \ .
\eeq

\noi That convergence will turn out to hold in the strong operator
topology. It implies in particular that for any bounded suitably
regular functions $R_j (a_0)$ $j = 1,2, \cdots , m$ and $R_j(\psi_0)$
$j= m+1, \cdots , \ell$ and for any family of times $\{t_j\}$ $j = 1,
2, \cdots , \ell$
$$\lim_{\lambda \to 0} C \left ( \alpha_{\lambda}(0)\right )^* \prod_{j=1}^m R_j \left ( a(t_j) - \alpha_{\lambda}(t_j) \right ) \prod_{j=m+1}^{\ell} R_j \left ( \psi (t_j)\right ) C \left ( \alpha_{\lambda}(0)\right )$$
$$= \prod_{j=1}^m R_j \left ( b'(t_j)\right ) \prod_{j=m+1}^{\ell} R_j \left ( \theta '(t_j)\right )$$

\noi in the strong sense. This convergence can be interpreted in terms
of correlation functions in coherent states of the Bose field. In
particular
$${\rm s}- \lim_{\lambda \to 0}  C \left ( \alpha_{\lambda}(0)\right )^* \prod_{j=1}^m R_j\left ( \lambda a (t_j)\right ) C \left ( \alpha_{\lambda}(0)\right ) = \prod_{j=1}^m R_j\left ( \alpha (t_j)\right )\ .$$

\noi In conclusion we expect that the weak coupling limit of the
quantum theory defined by the Hamiltonian $H$ (see (\ref{1.1e})
averaged over coherent states scaling as $\lambda^{-1}$ is the quantum
theory of nonrelativistic particles in an external potential $A$
solution of the equation
$$(\sq + \mu^2)A = 0\ .$$

\noi The function $A$ is the limit of the rescaled field $\lambda
\varphi$ and can be interpreted as the wave function of the condensate
of the excitations of the $\varphi$ field. The quantum fluctuations
around $A$ represent a free Bose field of mass $\mu$. \par

In the previous presentation we have totally ignored the fact that the
theory described by the Hamiltonian $H$ is ill defined. In order to
make $H$ a bonafide selfadjoint operator a cut off has to be
introduced. The removal of this cutoff, after subtraction of the
(infinite) self-energy of the nonrelativistic particles, has been
achieved by Nelson in \cite{8r} in the case $\mu > 0$ by using a
dressing transformation introduced by E.P. Gross \cite{5r}. The case
$\mu = 0$ has been subsequently treated by Fr\"ohlich \cite{2r}. Our aim
is to implement in a rigorous way the previously described limit when
$\lambda$ tends to zero and for that purpose we rely heavily on
\cite{8r}. For reasons of clarity the ideas behind the classical limit
have been so far explained by using the second quantization scheme for
the nonrelativistic particles. This has allowed us to treat the
particles and the field on the same footing with a similar formalism.
However, since the number of particles is conserved, we could have
worked as well in the first quantized formalism. In the next sections
we will follow this last option, namely we shall project the equations
(\ref{1.1e}), (\ref{1.22e}),  (\ref{1.24e}), (\ref{1.26e}) and
(\ref{1.27e}) on spaces with a fixed number of particles, keeping the
same notation for $H$, $W$, $V$ and $U$. \par

This paper is organized as follows. In Section 2, we recall without
proofs the results of Nelson that we need \cite{8r}. In Section 3 we
construct the limiting dynamics expressed by the propagator $V$. Finally
in Section 4 we prove the announced convergence when $\lambda \to 0$.
The main result is stated in Proposition 4.3 \par

We conclude this section by introducing some notation. We denote by
$\parallel\cdot \parallel_r$ the norm in $L^r \equiv L^r ({I\hskip-1truemm R}^{3})$, $1 \leq r
\leq \infty$, and by $(\cdot ,\cdot )$ the scalar product in $L^2$. We shall
need the spaces $L_s^2$ defined for any $s \in {I\hskip-1truemm R}$ by $L_s^2 = L_s^2({I\hskip-1truemm R}^{3}) =
\{ u:<\cdot >^s u \in L^2\}$ where $<\cdot > = (1 + |\cdot |^2)^{1/2}$.

\mysection{The quantum theory}
\hspace*{\parindent} In this section we describe the basic results
concerning the model we are interested in. We follow closely Nelson's
presentation and we refer essentially to \cite{8r} for the proofs. We
consider a system of $p$ nonrelativistic identical particles of mass
$M$ interacting with a real possibly massless Bose field. From now on
$p$ is fixed and generic constants $C$ in some of the subsequent
estimates may depend on $p$. The Hilbert space ${\cal H}$ of the
theory, which is the tensor product of $L^2({I\hskip-1truemm R}^{3p})$ and of the Fock space of
the Bose field, can be equivalently taken as the direct sum of the Hilbert spaces 
$${\cal H}_n = \{ \Psi_n : \Psi_n (x_1, \cdots , x_p;k_1, \cdots , k_n) \in L^2({I\hskip-1truemm R}^{3p + 3n})\}\ , \ n \geq 0$$

\noi with each $\Psi_n$ symmetric in the variables $k_1, \cdots , k_n$.
The set of variables $(x_1, x_2, \cdots ,x_p)$ is denoted by $X$. The
possible symmetry properties, if any, of the $\Psi_n$ in the $X$
variables do not play any role in the problem. The scalar product of
$\Phi$, $\Psi$ in ${\cal H}$ is denoted by $<\Phi , \Psi >$ and the
norm of $\Phi$ by $\parallel \Phi \parallel$. On ${\cal H}$ we define
formally the annihilation and creation operators for the Bose field by
$$\left ( a(k)\Psi\right )_n \left ( X;k_1, \cdots , k_n\right ) = (n+1)^{1/2} \Psi_{n+1}\left ( X;k,k_1, \cdots , k_n \right )\ ,$$
$$\left ( a^*(k)\Psi\right )_n \left ( X;k_1, \cdots , k_n\right ) = n^{-1/2} \sum_{j=1}^n \delta (k-k_j)\Psi_{n-1}\left ( X;k_1, \cdots , \widehat{k}_j, \cdots ,k_n \right )$$

\noi where $\widehat{k}_j$ indicates that the variable $k_j$ has been omitted. The field operator $\varphi$ is defined by 
$$\varphi (x) = (2\pi)^{-3/2} \int dk \ (2 \omega (k))^{-1/2} \left ( a(k) e^{ik\cdot x} + a^*(k) e^{-ik\cdot x} \right )\ . $$

\noi For any $f \in L^{\infty}({I\hskip-1truemm R}^{3p}, L^2({I\hskip-1truemm R}^{3}))$ we define (by formal integration) the operators
$$\left ( a(f)\Psi\right )_n \left ( X;k_1, \cdots , k_n\right ) = (n+1)^{1/2} \int dk\ f(X, k)\Psi_{n+1}\left ( X;k,k_1, \cdots , k_n \right )\ ,$$
$$\left ( a^*(f)\Psi\right )_n \left ( X;k_1, \cdots , k_n\right ) = n^{-1/2} \sum_{j=1}^n f (X, k_j)\Psi_{n-1}\left ( X;k_1, \cdots , \widehat{k}_j, \cdots ,k_n \right )\ .$$

\noi The number operator $N$, defined by
$$(N\Psi)_n \left ( X;k_1, \cdots , k_n\right ) = n \Psi_n \left ( X;k_1,  \cdots , k_n\right )\ ,$$

\noi counts the number of excitations of the Bose field. We denote by
${\cal C}_0 (N)$ the space of vectors in ${\cal H}$ with a finite number
of components different from zero. Standard estimates show that
$$\parallel a(f) \Psi \parallel \ \leq \ \parallel f;L^{\infty} ({I\hskip-1truemm R}^{3p},L^2({I\hskip-1truemm R}^{3}))\parallel \ \parallel N^{1/2} \Psi \parallel\ ,$$
$$\parallel a^*(f) \Psi \parallel \ \leq \ \parallel f;L^{\infty} ({I\hskip-1truemm R}^{3p},L^2({I\hskip-1truemm R}^{3}))\parallel \ \parallel (N +1)^{1/2} \Psi \parallel\ .$$

\noi For brevity from now on we shall write the estimates concerning
$a(f)$ and $a^*(f)$ as if $f$ did not depend on $X$, i.e. $f \in
L^2({I\hskip-1truemm R}^{3})$. The general case will be recovered by
replacing the norms of $f$ in $L^2({I\hskip-1truemm R}^{3})$ by the
corresponding norms in $L^{\infty}({I\hskip-1truemm R}^{3p},
L^2({I\hskip-1truemm R}^{3}))$ in the estimates. We now define the
dynamics of the theory. The kinetic energy $H_{01}$ of the
nonrelativistic particles is defined by
\beq
\label{2.1e}
\left ( H_{01} \Psi \right )_n \left ( X;k_1, \cdots , k_n\right ) = - (2M)^{-1} \sum_{j=1}^p \Delta_j \Psi_n \left ( X;k_1, \cdots , k_n\right )
\eeq

\noi where $\Delta_j$ is the Laplace operator acting on the $x_j$
variable, while the kinetic energy $H_{02}$ of the Bose field is
defined by 
\beq
\label{2.2e}
\left ( H_{02} \Psi \right )_n \left ( X;k_1, \cdots , k_n\right ) = \sum_{\ell =1}^n \omega (k_{\ell}) \Psi_n \left ( X;k_1, \cdots , k_n\right )
\eeq

\noi where $\omega (k) = (k^2 + \mu^2)^{1/2}$ and $\mu \geq 0$. We denote by $H_0$ their sum 
\beq
\label{2.3e}
H_0 = H_{01} + H_{02} \ .
\eeq

\noi It is well known that $H_0$ is self-adjoint on any ${\cal H}_n$
and therefore on the whole space ${\cal H}$. For any $\sigma$, $0 \leq
\sigma < \infty$, we define the cutoff function $\chi_{\sigma}$ by
$\chi_{\sigma}(k) = 1$ if $|k| \leq \sigma$, $\chi_{\sigma} (k) = 0$ if
$|k| > \sigma$. The interaction energy $H_{I\sigma}$ with cutoff $\sigma$
is defined by 
$$H_{I\sigma} = \lambda \sum_{j=1}^p \varphi_{\sigma} (x_j)$$

\noi with 
$$\varphi_{\sigma} (x) = (2 \pi)^{-3/2} \int dk \ (2 \omega (k))^{-1/2} \chi_{\sigma}(k) \left ( a(k) e^{ik\cdot x} + a^*(k) e^{-ik\cdot x}\right ) \ ,$$
\noi so that
\beq
\label{2.4e}
H_{I\sigma} = a\left ( \overline{f}\chi_{\sigma}\right ) + a^* \left ( f \chi_{\sigma}\right )
\eeq

\noi where
$$f = \sum_j f_j\ , \ f_j = f_0 \ e^{-ikx_j} \ (1 \leq j \leq p)\ ,$$
$$f_0 = \lambda (2 \pi)^{-3/2} (2 \omega (k))^{-1/2}\ .$$

\noi The sum of $H_0$ and $H_{I\sigma}$ defines the total Hamiltonian with cutoff
\beq
\label{2.5e}
H_{\sigma} = H_0 + H_{I\sigma}\ .
\eeq

\noi If we take formally $\sigma = \infty$, namely $\chi_{\sigma}(k) =
1$, the second quantized version of the expression given by
(\ref{2.5e}) coincides with the expression (\ref{1.1e}). \par

The analysis of Nelson and Fr\"ohlich is based on some estimates of
$a(f)$ and $a^*(f)$ in terms of the operator $H_0$. The following set
of estimates holds for all $\mu \geq 0$. \\

\noi {\bf Lemma 2.1.} {\it Let $f \in L^2({I\hskip-1truemm R}^{3})$ with
$\omega^{-1/2} f \in L^2({I\hskip-1truemm R}^{3})$. Then, for all
$\Psi$, $\Phi \in {\cal H}$, the following estimates hold~:}
\beq
\label{2.6e}
\parallel a(f)\Psi \parallel^2 \ \leq \ \parallel \omega^{-1/2} f \parallel_2^2 \ \parallel H_{02}^{1/2} \Psi \parallel^2 \ ,
\eeq
\beq
\label{2.7e}
\parallel a^*(f)\Psi \parallel^2 \ \leq \ \parallel \omega^{-1/2} f \parallel_2^2 \ \parallel H_{02}^{1/2} \Psi \parallel^2 \ + \ \parallel f\parallel_2^2 \ \parallel \Psi \parallel_2^2 \ ,
\eeq
\beq
\label{2.8e}
\parallel a(f)^2\Psi \parallel^2 \ \leq \ \parallel \omega^{-1/2} f \parallel_2^4 \ \parallel H_{02} \Psi \parallel^2 \ ,
\eeq
\beq
\label{2.9e}
\parallel a^*(f) a(f)\Psi\parallel^2 \ \leq \ \parallel \omega^{-1/2} f \parallel_2^4 \ \parallel H_{02} \Psi \parallel^2 \ + \ \parallel f\parallel_2^2 \ \parallel \omega^{-1/2} f \parallel_2^2 \ \parallel H_{02}^{1/2} \Psi \parallel^2\ ,
\eeq
\bea
\label{2.10e}
\parallel \left ( a^*(f)\right )^2 \Psi\parallel^2 &\leq & \parallel \omega^{-1/2} f \parallel_2^4 \ \parallel H_{02} \Psi \parallel^2 \ + \ 4\parallel f\parallel_2^2 \ \parallel \omega^{-1/2} f \parallel_2^2 \ \parallel H_{02}^{1/2} \Psi \parallel^2\nn \\
&&+ 2 \parallel f\parallel_2^4 \ \parallel \Psi \parallel^2 \ ,
\eea
\bea
\label{2.11e}
\left | <\Phi , \left ( a(f)\right )^2 \Psi > \right |& \leq &(3/2)^{1/2} \Big ( \parallel \omega^{-1/2}f\parallel_2^2 \ \parallel H_{02}^{1/2}\Psi \parallel \  \parallel H_{02}^{1/2} \Phi \parallel \nn \\
&&+ \parallel \omega^{-1/4}f\parallel_2^2 \ \parallel H_{02}^{1/2}\Psi \parallel \ \parallel \Phi \parallel \Big )\ .
\eea

\noi Using (\ref{2.6e}) and (\ref{2.7e}) we obtain the following result.\\

\noi {\bf Proposition 2.1.} {\it For any $\sigma < \infty$, the operator $H_{\sigma}$ is self-adjoint on ${\cal D} (H_0)$.}\\

In order to remove the cut off $\sigma$, we use a dressing
transformation which allows to change the domain of definition of the
limiting Hamiltonian with respect to the domain of $H_0$. In addition
to the upper cut off $\sigma$ we introduce a lower cut off $\sigma_0 <
\sigma$ which we keep fixed and which eventually will be chosen
sufficiently large. In analogy with (\ref{2.4e}), we define the
operators 
\beq
\label{2.12e}
T_{\sigma} = a \left ( \overline{g} \chi_{\sigma}\right ) - a^* \left ( g \chi_{\sigma}\right ) \ ,
\eeq
\beq
\label{2.13e}
T = a (\overline{g}) - a^*(g) \ ,
\eeq

\noi where
$$g = \sum_j g_j \ , \ g_j = g_0 \ e^{-ikx_j} \ (1 \leq j \leq p ) \ ,$$
\begin{eqnarray*}
g_0(k) &=& - \left ( 1 - \chi_{\sigma_0}(k)\right ) \left ( \omega (k) + (2M)^{-1}k^2\right )^{-1} \ f_0(k)\\
&=& - \left ( 1 - \chi_{\sigma_0}(k)\right ) \lambda \left ( \omega (k) + (2M)^{-1}k^2\right )^{-1} \ (2 \pi)^{-3/2} (2 \omega (k))^{-1/2} \ .
\end{eqnarray*}

\noi Note in particular that $g_0$ and therefore also $g_j$ and $g$
belong to $L^2$. We identify the operators $T_{\sigma}$ and $T$ with
their closures. We also define the operators
\beq
\label{2.14e}
Q_{\sigma} = \exp \left ( - T_{\sigma}\right ) \quad , \quad Q = \exp (-T)\ .
\eeq

\noi The operators $Q_{\sigma}$ and $Q$ are Weyl operators associated with the
Bose field $(a, a^*)$. In addition, they depend on the coordinates $X$,
and therefore they also act as operators in the tensor factor
$L^2({I\hskip-1truemm R}^{3p})$ of ${\cal H}$.\par

The operators $T_{\sigma}$, $T$ and $Q_{\sigma}$, $Q$ enjoy the following properties.\\

\noi {\bf Proposition 2.2.} \par
{\it 1) The operators $i T_{\sigma}$ and $iT$ are essentially self-adjoint
on ${\cal C}_0 (N)$. The operators $Q_{\sigma}$ and $Q$ are unitary.
\par

2) $Q_{\sigma}{\cal D} (H_0^{1/2}) = {\cal D}(H_0^{1/2})$ and $Q_{\sigma}
{\cal D}(H_0) = {\cal D} (H_0)$. \par

3) The following limit holds in the strong sense}
$${\rm s}- \lim_{\sigma \to \infty} Q_{\sigma} = Q\ .$$
\vskip 5 truemm

Upon formal transformation of $H_{\sigma}$ by the unitary operator $Q_{\sigma}$, we obtain for $\sigma_0 < \sigma$
\beq
\label{2.15e}
\left ( H_{\sigma} - p E_{\sigma}\right ) Q_{\sigma} = Q_{\sigma} \left ( H'_{\sigma} - p E_{\sigma_0}\right )
\eeq
\noi where
\beq
\label{2.16e}
E_{\sigma} = - \lambda^2 (2 \pi)^{-3} \int dk \ (2 \omega (k))^{-1} \left ( \omega (k) + (2M)^{-1} k^2\right )^{-1} \chi_{\sigma}(k) \ ,
\eeq 
\beq
\label{2.17e}
H'_{\sigma} = H_0 + H_{I\sigma_0} + H'_{1\sigma} + H'_{2\sigma} + H'_{3\sigma}\ ,
\eeq
\beq
\label{2.18e}
H'_{1\sigma} = i M^{-1} \sum_{j=1}^p \left \{ \nabla_j \cdot a \left ( k \overline{g}_j \chi_{\sigma}\right ) + a^* \left ( k g_j \chi_{\sigma}\right ) \cdot \nabla_j \right \} \ ,
\eeq
\beq
\label{2.19e}
H'_{2\sigma} =  (2M)^{-1} \sum_{j=1}^p \left \{ \left ( a \left ( k \overline{g}_j \chi_{\sigma}\right )\right )^2 + \left ( a^* \left ( k g_j \chi_{\sigma}\right )\right )^2 + 2a^* \left ( k g_j \chi_{\sigma}\right ) a \left ( k \overline{g}_j \chi_{\sigma}\right )  \right \} \ ,
\eeq
\beq
\label{2.20e}
H'_{3\sigma} = \sum_{1 \leq j < \ell \leq p} q_{\sigma} \left (  x_j - x_{\ell} \right ) \ ,
\eeq
\bea
\label{2.21e}
q_{\sigma}(x) &=& - 2 \lambda^2 (2 \pi)^{-3} \int dk \ (2 \omega (k))^{-1} \left ( \omega (k) + (2M)^{-1} k^2\right )^{-2} \left ( \omega (k) + M^{-1} k^2 \right )\nn \\
&&\times \left ( \chi_{\sigma} (k) - \chi_{\sigma_0} (k)\right ) \cos k x \ .
\eea

\noi The following proposition summarizes the meaning of the equality (\ref{2.15e}).\\

\noi {\bf Proposition 2.3.} {\it Let $0 < \sigma_0 < \sigma$. Then\par

1) The operator $H'_{\sigma}$ is self-adjoint on ${\cal D} (H_0)$. \par
2) The equality (\ref{2.15e}) holds on ${\cal D}(H_0)$.}\\

In order to remove the cutoff $\sigma$ and to take later the limit
$\lambda \to 0$, the estimates contained in the following lemma will be
useful. They are a consequence of Lemma 2.1.\\

\noi {\bf Lemma 2.2.} {\it For all $\sigma$ with $0 < \sigma_0 < \sigma$, for all $\Psi \in
{\cal D} (H_0^{1/2})$, the following estimates hold~:}
$$\left | < \Psi, H_{I\sigma_0}\Psi > \right | \leq \lambda \left ( \varepsilon \parallel H_0^{1/2} \Psi \parallel^2 \ + \ C\ \varepsilon^{-1}\ \sigma_0 \parallel \Psi \parallel^2 \right ) \ ,$$
$$\left | < \Psi, H'_{1\sigma}\Psi > \right | \leq C\lambda \ \sigma_0^{-1/2} \parallel H_0^{1/2} \Psi \parallel^2\ ,$$ 
$$\left | < \Psi, H'_{2\sigma}\Psi > \right | \leq C\lambda^2 \left (  \sigma_0^{-1} \parallel H_0^{1/2} \Psi \parallel^2\ + \ \parallel \Psi \parallel^2 \right )\ ,$$ 
$$\left | < \Psi, H'_{3\sigma}\Psi > \right | \leq \lambda^2 \left (  \varepsilon \parallel H_0^{1/2} \Psi \parallel^2\ + \ C\ \varepsilon^{-1} \ \sigma_0^{-2} \parallel \Psi \parallel^2 \right )\ .$$ 

Finally the above estimates and similar ones lead to the existence of a
limiting operator $H'_{\infty} = \displaystyle{\lim_{\sigma \to
\infty}} H'_{\sigma}$ in the sense of the form defined by the operator
$1 + H_0$ and therefore to the existence of the renormalized
Hamiltonian of the theory $\widehat{H}$ defined by 
\beq
\label{2.22e}
\widehat{H} = Q\left ( H'_{\infty} - p E_{\sigma_0}\right ) Q^*
\eeq
\noi (see (\ref{2.15e})). \\

\noi {\bf Proposition 2.4.} {\it Let $\lambda_0 > 0$. Then there exists $\sigma_0 > 0$ such that
for all $\lambda$, $|\lambda | \leq \lambda_0$,\par 1) For all $\sigma$
with $\sigma_0 < \sigma$ , for all $\Psi \in {\cal D} (H_0^{1/2})$, the
following estimate holds 
$$\left | < \Psi, (H'_{\sigma}- H_0)\Psi > \right | \leq (1/2) \parallel H_0^{1/2} \Psi \parallel^2\ + \ C \parallel \Psi \parallel^2\ .$$

2) For all $\sigma_1$, $\sigma_2$ with $\sigma_0 < \sigma_1 < \sigma_2$
, for all $\Psi \in {\cal D} (H_0^{1/2})$, the following estimate holds 
$$\left | < \Psi, (H'_{\sigma_1}- H'_{\sigma_2})\Psi > \right | \leq \varepsilon (\sigma_1) \left ( \parallel H_0^{1/2} \Psi \parallel^2\ + \ \parallel \Psi \parallel^2\right ) \ ,$$

\noi where $\varepsilon (\sigma_1)$ tends to zero when $\sigma_1$ tends to infinity. \par

3) There exists a self-adjoint operator $H'_{\infty}$ such that for all
$\sigma$ with $\sigma_0 < \sigma$ and for all $\Psi \in {\cal
D}(H_0^{1/2})$ the following estimate holds
$$\left | < \Psi, (H'_{\sigma}- H'_{\infty})\Psi > \right | \leq \varepsilon (\sigma ) \left ( \parallel H_0^{1/2} \Psi \parallel^2\ + \ \parallel \Psi \parallel^2\right ) \ .$$

\noi The operator $H'_{\infty}$ is bounded from below and there exist two constants $\rho$ and $C$ such that 
$$C^{-1} \parallel (1 + H_0)^{1/2} \Psi\parallel^2\ \leq \ <\Psi, (\rho + H'_{\infty} )\Psi >\ \leq \ C \parallel (1 + H_0)^{1/2} \Psi \parallel^2$$

\noi for all $\Psi \in {\cal D} (H_0^{1/2})$. Furthermore
\beq
\label{2.23e}
{\rm s} - \lim_{\sigma \to \infty} \exp \left ( -itH'_{\sigma} \right ) = \exp \left ( - i t H'_{\infty}\right ) \ .
\eeq

\noi The operator $\widehat{H}$ is selfadjoint and
\beq
\label{2.24e}
{\rm s} - \lim_{\sigma \to \infty} U_{\sigma}(t) = U(t) \equiv \exp \left ( - i t \widehat{H}\right ) 
\eeq

\noi where
\beq
\label{2.25e}
U_{\sigma}(t)\equiv  \exp \left \{ -it\left ( H_{\sigma} - p E_{\sigma} \right )\right \} \ .
\eeq

\noi Both limits (\ref{2.23e}) and (\ref{2.24e}) hold for any $t \in {I\hskip-1truemm R}$ uniformly on compact intervals.}

\mysection{The limiting theory}
\hspace*{\parindent} In this section we give a precise definition and we study the
properties of the unitary propagator implicitly and formally defined by
(\ref{1.24e}) in the second order formalism. Rewritten in the first
order formalism the problem consists in solving the equation
 \beq
\label{3.1e}
\left \{ \begin{array}{l} i \partial_t V(t, s) = \left ( H_0 + {\cal A}(t)\right ) V(t,s)\\ \\ V(s,s) = \1\end{array} \right .
\eeq

\noi where
\beq
\label{3.2e}
\left ( {\cal A}(t)\Psi\right )_n \left ( X; k_1, \cdots , k_n\right ) = \sum_{j=1}^p A(t, x_j) \Psi_n \left ( X;k_1, \cdots , k_n\right ) \ ,
\eeq
\beq
\label{3.3e}
A(t,x) = (2 \pi)^{-3/2}\int dk \ (2 \omega (k))^{-1/2} \left ( \alpha (k) e^{i(k\cdot x - \omega (k)t)} + \overline{\alpha (k)} \ e^{-i(k\cdot x - \omega (k) t)}\right )\ .
\eeq

\noi The function $A$ defined by (\ref{3.3e}) is the function defined
by (\ref{1.9e}) with $\alpha (t, k) = \alpha (k) \exp (-i\omega (k)
t)$, namely with $\alpha (t)$ solution of (\ref{1.11e}). We define in
addition 
\beq
\label{3.4e}
\left ( \dot{\cal A}(t)\Psi\right )_n \left ( X; k_1, \cdots , k_n\right ) = \sum_{j=1}^p (\partial_t A)(t, x_j) \Psi_n \left ( X;k_1, \cdots ,k_n\right ) 
\eeq

\noi which represents the time derivative of the family of operators
${\cal A}(t)$. We collect some properties of ${\cal A}$ in the next
lemma. \\

\noi {\bf Lemma 3.1.} {\it Let $\alpha \in L_{1/2}^2$. Then the
operators ${\cal A}(t)$ and $\dot{\cal A}(t)$ satisfy the following
estimates
\beq
\label{3.5e}
\parallel {\cal A}(t) \Psi \parallel \ \leq \ C \parallel \omega^{1/2} \alpha \parallel_2\ \parallel H_{01}^{1/4} \Psi \parallel \ ,
\eeq
\beq
\label{3.6e}
\left | < \Psi, {\cal A}(t) \Psi >\right | \ \leq \ C \parallel \omega^{1/2} \alpha \parallel_2 \ < \Psi, H_{01}^{1/4} \Psi> \ ,
\eeq
\beq
\label{3.7e}
\parallel \dot{\cal A}(t) \Psi \parallel \ \leq \ C \parallel \omega^{1/2} \alpha \parallel_2\ \parallel \Psi \parallel^{1/4}\ \parallel H_{01}\Psi \parallel^{3/4}  \ ,
\eeq
\beq
\label{3.8e}
\left | < \Psi, \dot{\cal A}(t) \Psi >\right | \ \leq \ C \parallel \omega^{1/2} \alpha \parallel_2 \ < \Psi, H_{01}^{3/4} \Psi > \ ,
\eeq
\beq
\label{3.9e}
\parallel \left ( (t-s)^{-1} \left ( {\cal A}(t) - {\cal A}(s)\right ) - \dot{\cal A}(s)\right ) \Psi \parallel \ \leq \ C \parallel \omega^{1/2} \alpha \ m(s,t)\parallel_2\ \parallel \Psi \parallel^{1/4}\ \parallel H_{01} \Psi \parallel^{3/4} 
\eeq

\noi with}
$$m(s,t) = 1 - \int_0^1 e^{i\omega (s-t)\theta} \ d\theta\ .$$
\vskip 5 truemm

\noi {\bf Proof.} From the definition of $A(t)$ and $(\partial_t A) (t)$ we obtain 
\beq
\label{3.10e}
\parallel A(t)\parallel_6 \ \leq \ C \parallel \omega^{1/2} \alpha \parallel_2
\eeq

\noi by a Sobolev inequality and 
\beq
\label{3.11e}
\parallel \partial_t A(t) \parallel_2 \ \leq \ C \parallel \omega^{1/2} \alpha \parallel_2
\eeq

\noi by the unitarity of the Fourier transform. \par

We first prove (\ref{3.5e}). Let $X = (x_1, X')$. Then 
$$\int dx_1 \left | A(t, x_1) \Psi_n \left ( x_1, X';k_1, \cdots , k_n\right ) \right |^2 \leq\ \parallel A(t)\parallel_6^2 \ \parallel \Psi_n \left (\cdot, X';k_1, \cdots , k_n \right )\parallel_3^2$$
$$\leq\ C \parallel A(t)\parallel_6^2\ \parallel \Delta_1^{1/4} \Psi_n \left ( \cdot, X';k_1, \cdots , k_n \right )\parallel_2^2$$

\noi by H\"older and Sobolev inequalities. Integrating over the variables $X'$ and $k_1, \cdot , k_n$ and summing over $n$ we obtain
$$\parallel {\cal A}(t) \Psi\parallel \ \leq \ C \parallel A(t) \parallel_6 \sum_{j=1}^p \parallel \Delta_j^{1/4} \Psi \parallel$$

\noi which implies (\ref{3.5e}) by (\ref{3.10e}). The proof of (\ref{3.6e}) is similar.\par

We next prove (\ref{3.7e}). We estimate
$$\int dx_1 \left | (\partial_t A)(t, x_1) \Psi_n \left ( x_1, X';k_1, \cdots ,k_n\right ) \right |^2 \leq\ \parallel \partial_t A(t)\parallel_2^2 \ \parallel \Psi_n \left ( \cdot, X';k_1, \cdots ,k_n \right )\parallel_{\infty}^2$$
$$\leq\ C \parallel (\partial_t A)(t)\parallel_2^2\ \parallel \Psi_n \left ( \cdot, X';k_1, \cdots ,k_n \right )\parallel_2^{1/2} \ \ \parallel \Delta_1\Psi_n \left ( \cdot, X';k_1, \cdots ,k_n \right )\parallel_2^{3/2}$$

\noi by H\"older and Sobolev inequalities. Integrating over the variables $X'$ and $k_1 , \cdots , k_n$ and summing over $n$ we obtain 
$$\parallel \dot{\cal A}(t) \Psi\parallel \ \leq \ C \parallel (\partial_t A)(t) \parallel_2 \ \parallel \Psi \parallel^{1/4} \sum_{j=1}^p \parallel \Delta_j \Psi \parallel^{3/4}$$

\noi which implies (\ref{3.7e}) by (\ref{3.11e}). The proof of
(\ref{3.8e}) is similar. We finally prove (\ref{3.9e}). By application
of (\ref{3.2e}), (\ref{3.3e}) and (\ref{3.4e}) we can write
$$\left ( \left ( (t-s)^{-1} ({\cal A}(t) - {\cal A}(s)) - \dot{\cal A}(s) \right ) \Psi \right )_n \left ( X;k_1 , \cdots , k_n\right ) =$$
$$= \sum_{j=1}^p  B(s,t,x_j) \Psi_n \left ( X;k_1 , \cdots , k_n\right )$$

\noi where
$$B(s,t,x) = i (2\pi)^{-3/2}\int dk \ (2 \omega (k))^{-1/2} \omega (k) \Big \{ \alpha (k) \ e^{i(kx-\omega (k) s)} m(s,t)$$
$$- \overline{\alpha (k)} \ e^{-i(k\cdot x - \omega (k) s)}\ \overline{m(s,t)} \Big \}$$

\noi Now the remaining part of the proof is identical with that of
(\ref{3.7e}). \par \nobreak \hfill $\sq$ \par

We are now in condition to prove the existence and uniqueness of
solutions $V(t,s)$ of (\ref{3.1e}). For that purpose we rely on a
result of Kato \cite{7r}.\\

\noi {\bf Proposition 3.1.} {\it Let $\alpha \in L_{1/2}^2$ and let
${\cal A}(t)$ be defined by (\ref{3.2e}) and (\ref{3.3e}). Then \par

1) For any $t\in {I\hskip-1truemm R}$, ${\cal A}(t)$ is a Kato perturbation of $H_0$, so that $H_0 + {\cal A}(t)$ is self-adjoint on ${\cal D}(H_0)$.\par

2) There exists a family of unitary operators $V(t,s)$, $t, s \in {I\hskip-1truemm R}$ with the following properties\par

(a) $V(t,t) = \1$.\par
(b) $V(t,s)\ V(s,r) = V(t,r)$.\par
(c) $V(t,s)$ is strongly continuous on ${I\hskip-1truemm R} \times {I\hskip-1truemm R}$.\par
(d) $V(t,s) \ {\cal D}(H_0) \subset {\cal D}(H_0)$\par

\noi and for any compact interval $I$ there exists a constant $C_I$ such that
\beq
\label{3.12e}
\parallel (1 + H_0) V(t,s) \Psi \parallel\ \leq \ C_I \parallel (1 + H_0)\Psi \parallel
\eeq

\noi for any $\psi \in {\cal D}(H_0)$ and for all $t$, $s \in I$. \par
(e) For any $\Psi \in {\cal D}(H_0)$
$$i\ {d \over dt} V(t,s) \Psi = \left ( H_0 + {\cal A}(t)\right ) V(t, s) \Psi\ .$$

3) Uniqueness holds under the assumptions (a), (d) and (e).}\\

\noi {\bf Proof.} \par

1) From (\ref{3.5e}) it follows that for any $t \in {I\hskip-1truemm
R}$ and for any $\Psi \in D(H_0)$ the following inequality holds~: 
$$\parallel {\cal A}(t)\Psi \parallel \ \leq \ \varepsilon \parallel H_{01}\Psi \parallel \ + \ C \varepsilon^{-1/3} \parallel \omega^{1/2}\alpha \parallel_2^{4/3} \ \parallel \Psi \parallel$$

\noi so that ${\cal A}(t)$ is infinitesimally small with respect to
$H_0$. Therefore, for any $t \in {I\hskip-1truemm R}$, $H_0 + {\cal
A}(t)$ is self-adjoint on ${\cal D}(H_0)$. \par

2) The existence of $V(t, s)$ and its properties follows from Theorem 1
of \cite{7r} once we have verified the assumptions of the theorem. The
only non trivial point consists in proving that for some $\rho$ and for
any $t\in {I\hskip-1truemm R}$ the operator
$$S(t) \equiv \rho + H_0 + {\cal A}(t)$$

\noi is an isomorphism of ${\cal D}(H_0)$ onto ${\cal H}$ and that, for
any $\Psi \in {\cal D}(H_0)$, $S(t)\Psi$ is continuously differentiable.
From (\ref{3.5e}) it follows that there exist $\rho$ and $C$ such that
$$C^{-1} \parallel (1 + H_0)\Psi \parallel \ \leq \ \parallel (\rho + H_0 + {\cal A}(t))\Psi \parallel\ \leq \ C\parallel (1 + H_0)\Psi\parallel$$   

\noi for any $\Psi \in {\cal D}(H_0)$ and for any $t\in {I\hskip-1truemm R}$. This leads to the isomorphism property. Since 
$$S(t) - S(s) = {\cal A}(t) - {\cal A}(s)$$

\noi the differentiability properties of $S$ are the differentiability
property of ${\cal A}$. By (\ref{3.9e}) we see that $S(t)\Psi$ is
differentiable and that
$${d \over dt}\ S(t) \Psi = \dot{\cal A} (t) \Psi \ .$$

\noi The continuity of $\dot{\cal A}(t)\Psi$ follows from a minor variation of (\ref{3.7e}). \par

3) To prove uniqueness let us suppose the existence of $V'(t,s)$
satisfying (a), (d) and (e). Then, for any $\Phi$, $\Psi \in {\cal
D}(H_0)$ the conditions (d) and (e) imply 
$${d \over dt} \ <V'(t,s)\Phi, V(t, s)\Psi > \ = 0$$

\noi where $V(t,s)$ is the family constructed in Part 2. On the other hand
$${d\over dt} \ <V(t,s)\Phi, V(t,s)\Psi > \ = 0$$

\noi so that by the condition (a)
$$<V'(t,s)\Phi - V(t,s)\Phi , V(t,s)\Psi > \ = 0$$

\noi which implies
$$V'(t,s)\Phi = V(t,s) \Phi \ .$$
\par \nobreak \hfill $\sq$\par

\mysection{The limit $\lambda \to 0$}
\hspace*{\parindent}In this section we prove the main result of this paper, namely the
operator convergence when $\lambda \to 0$ announced in (\ref{1.27e}).
As in Section 2, we use the first order formalism for the particles. We
recall the definition of the Weyl operator (see (\ref{1.8e}) where it
is written with $a_0$ instead of $a$)

\beq
\label{4.1e}
C(\alpha ) = \exp \left ( a^*(\alpha ) - a(\overline{\alpha})\right )
\eeq

\noi for any $\alpha \in L^2$. (Strictly speaking the operator in the
exponential should be replaced by its closure). We now define (see
(\ref{1.25e}))
\beq
\label{4.2e}
W(t,s) = C \left ( \alpha_{\lambda}(t)\right )^* U(t-s) C \left ( \alpha_{\lambda}(s)\right )
\eeq

\noi where
\beq
\label{4.3e}
\alpha (t, k) = \alpha (k) \exp (- i \omega t)\ ,
\eeq

\noi $\alpha_{\lambda} = \lambda^{-1} \alpha$ and $U$ is defined in (\ref{2.24e}).
Although $U$ and $W$ depend on $\lambda$, for brevity we shall omit
that dependence. We intend to prove that $W(t,s)$ converges strongly
when $\lambda \to 0$ to the propagator $V(t,s)$ defined in Proposition
(3.1), uniformly for $t,s$ in compact intervals. The following lemma
collects some properties of the Weyl operators.\\

\noi {\bf Lemma 4.1.} \par

{\it 1) Let $\alpha \in L^2$. Then $C(\alpha )$ is unitary and strongly
continuous as a function of $\alpha \in L^2$. In addition, for any $\Psi
\in {\cal D} (a (\overline{\gamma}))$ with $\gamma \in L^2$, $C(\alpha )
\Psi \in {\cal D} (a(\overline{\gamma}))$ and the following identity
holds~:
\beq
\label{4.4e}
C(\alpha )^* a(\overline{\gamma}) C(\alpha ) \Psi = a (\overline{\gamma}) \Psi + (\gamma, \alpha ) \Psi \ .
\eeq

\noi Similarly, for any $\Psi \in {\cal D} (a^*(\gamma ))$, $C(\alpha )
\Psi \in {\cal D} (a^*(\gamma ))$ and the following identity holds~: 
\beq
\label{4.5e}
C(\alpha )^* a^*(\gamma ) C(\alpha ) \Psi = a^* (\gamma) \Psi + (\alpha , \gamma) \Psi \ .
\eeq

2) Let $\alpha \in L_{1/2}^2$. Then $C(\alpha ) {\cal D}(H_0^{1/2}) =
{\cal D}(H_0^{1/2})$ and, for any $\Psi \in {\cal D} (H_0^{1/2})$, the
following inequality holds~:
\beq
\label{4.6e}
\parallel H_0^{1/2} C( \alpha ) \Psi \parallel \ \leq \ \parallel H_0^{1/2} \Psi \parallel \ + \ \parallel \omega^{1/2} \alpha \parallel\ \parallel \Psi \parallel \ .
\eeq

\noi Let $\alpha \in L_1^2$. Then $C(\alpha ) {\cal D} (H_0) = {\cal
D}(H_0)$ and, for any $\Psi \in {\cal D} (H_0)$, the following
inequality holds~:
\beq
\label{4.7e}
\parallel H_0 C( \alpha ) \Psi \parallel \ \leq \ 2\parallel H_0 \Psi \parallel \ + \ \left ( \parallel \omega \alpha \parallel\ +\ 2\parallel \omega^{1/2} \alpha \parallel^2\right ) \parallel \Psi \parallel \ .
\eeq

3) Let $\alpha : t \to \alpha (t) \in {\cal C}^1 ({I\hskip-1truemm R},
L^2)$ with $\omega^{-1/2} d\alpha/dt \equiv \omega^{-1/2} \dot{\alpha} \in {\cal C}({I\hskip-1truemm R},
L^2)$. Then, for any $\Psi \in {\cal D} (H_0^{1/2})$, $C(\alpha
(t))\Psi$ is differentiable in $t$. Its derivative is given by}
\beq
\label{4.8e}
{d \over dt} C(\alpha (t))\Psi = C(\alpha (t)) \left ( a^* (\dot{\alpha}(t)) - a (\overline{\dot{\alpha}(t)}) + i \ {\rm Im} (\alpha (t), \dot{\alpha} (t))\right ) \Psi\ .
\eeq

\noi {\bf Proof.}\par
1) The set of vectors ${\cal C}_0 (N)$ is a domain of essential
self-adjointness for $i(a^*(\alpha ) - a (\overline{\alpha}))$ so that
$C(\alpha )$ is unitary. In addition ${\cal C}_0 (N)$ is a set of
entire analytic vectors \cite{10r} for $a^*(\alpha ) -
a(\overline{\alpha })$, which leads to the continuity of $C(\alpha
)\Psi$ in $\alpha$ for any $\Psi \in {\cal C}_0 (N)$ by direct
inspection. Strong continuity for any $\Psi$ follows immediately. Using
again the power series expansion of $C(\alpha )\Psi$ for $\Psi \in {\cal
C}_0(N)$ we can check immediately that (\ref{4.4e}) holds for such a
$\Psi$. An elementary argument of closure leads to (\ref{4.4e}) in
general. The proof of the part concerning $a^*(\gamma )$ is similar.\par

2) Let $\alpha \in L_1^2$. By power series expansion we check directly
that, for any $\Psi \in {\cal C}_0 (N) \cap {\cal D} (H_{02})$, $C
(\alpha ) \Psi \in {\cal D}(H_{02})$ and that the following identity
holds~:
\beq
\label{4.9e}
H_{02} C( \alpha ) \Psi = C(\alpha )  \left (  H_{02}  + a^*(\omega \alpha ) + a(\omega \overline{\alpha}) + \parallel \omega^{1/2} \alpha \parallel^2 \right ) \Psi  \ .
\eeq
 
\noi Using (\ref{2.6e}) and (\ref{2.7e}) with $f = \omega \alpha$ and
the Schwarz inequality, we obtain (\ref{4.7e}). A standard
approximation argument leads to the conclusion that $C(\alpha ) {\cal
D}(H_{02}) \subset {\cal D} (H_{02})$ and that (\ref{4.7e}) holds for
any $\Psi \in {\cal D}(H_0)$.\par

Similarly from (\ref{4.9e}), using (\ref{2.6e}), we obtain (\ref{4.6e})
for $\Psi \in {\cal C}_0(N) \cap {\cal D}(H_{02})$. To conclude we
apply an approximation argument first on $\Psi$ and then on
$\alpha$.\par

3) The Weyl operators satisfy the following well known identity 
\beq
\label{4.10e}
C(\alpha + \beta) = C(\alpha ) C(\beta ) \exp \left ( i\ {\rm Im} (\alpha , \beta )\right )
\eeq

\noi which can be proved by power series expansion on ${\cal C}_0 (N)$
and then extended to the whole Hilbert space ${\cal H}$ by the
unitarity of $C(\alpha )$. Using (\ref{4.10e}) applied to $\Psi \in {\cal C}_0
(N)$, we can write the identity
\bea
\label{4.11e}
&&(t-t_0)^{-1} \left ( C(\alpha (t)) - C(\alpha (t_0))\right ) \Psi = C(\alpha (t_0)) (t-t_0)^{-1}\nn \\
&&\times \left \{ C\left ( \alpha (t) - \alpha (t_0)\right ) \exp \left ( i\ {\rm Im} (\alpha (t_0), \alpha (t) - \alpha (t_0))\right ) - 1 \right \} \Psi 
\eea

\noi which in the limit $t \to t_0$ yields (\ref{4.8e}). Let now $\Psi
\in {\cal D} (H_{02}^{1/2})$. We write the integrated form of
(\ref{4.8e}), namely 
\bea
\label{4.12e}
&& C(\alpha (t))\Psi_j = C(\alpha (t_0)) \Psi_j + \int_{t_0}^t ds\  C(\alpha (s)) \nn \\
&&\times \left ( a^* ( \dot{\alpha} (s)) - a (\overline{\dot{\alpha}(s)}) +  i\ {\rm Im} (\alpha (s), \dot{\alpha} (s) )\right ) \Psi_j 
\eea

\noi for a sequence $\Psi_j \in {\cal C}_0 (N) \cap {\cal D}
(H_{02}^{1/2})$ such that $\Psi_j \to \Psi$ and $H_{02}^{1/2}\Psi_j \to
H_{02}^{1/2}\Psi$. Using (\ref{2.6e}) and (\ref{2.7e}) we can take
the limit $j \to \infty$ in both sides of (\ref{4.12e}) and we obtain
(\ref{4.12e}) with $\Psi_j$ replaced by $\Psi$. By differentiation in
$t$ we obtain (\ref{4.8e}) in full generality. \par \nobreak \hfill
$\sq$ \par

We continue the argument temporarily with the approximate theory
defined by the Hamiltonian $H_{\sigma} - p E_{\sigma}$ (see Prop. 2.4,
part 3) and we shall remove the cutoff $\sigma$ at the end. For that
purpose, for any $\sigma$ with $\sigma_0 < \sigma$ we define
\beq
\label{4.13e}
Z_{\sigma}(t,s) = Q_{\sigma}^*\ C\left ( \chi_{\sigma}   \alpha_{\lambda} (t)\right )^* \ U_{\sigma} (t-s) \ C \left ( \chi_{\sigma}   \alpha_{\lambda} (s) \right ) \ Q_{\sigma}
\eeq 

\noi where $Q_{\sigma}$ and $U_{\sigma}$ are defined by (\ref{2.14e})
and (\ref{2.25e}) respectively, and $\alpha_{\lambda} (t) =
\lambda^{-1} \alpha (t)$ with $\alpha$ given by (\ref{4.3e}). In
addition we define ${\cal A}_{\sigma} (t)$ by (\ref{3.2e}) with $A$
replaced by $A_{\sigma}$ where 
\beq
\label{4.14e}
A_{\sigma}(t) = (2 \pi)^{-3/2} \int dk \ (2 \omega (k))^{-1/2} \chi_{\sigma}(k) \left ( \alpha (k) e^{i(kx-\omega (k)t)} + \overline{\alpha (k)} e^{-i(kx - \omega (k)t)}\right )\ .
\eeq

In the following proposition we perform the basic computation which
exhibits the compensations among the terms containing the coupling
constant $\lambda$ in the operator $Z_{\sigma}(t,s)$. \\

\noi {\bf Proposition 4.1.} {\it  Let $\alpha \in L_{1/2}^2$ and let
$\alpha (t)$ be given by (\ref{4.3e}). Let $\Psi \in {\cal D}(H_0)$.
Then, for any $\sigma$ with $\sigma_0 < \sigma$, $Z_{\sigma} (t,
s)\Psi$ is differentiable in $t$ with derivative given by 
\beq
\label{4.15e}
i\ {d \over dt} \ Z_{\sigma} (t, s) \Psi = \left ( H'_{\sigma} - p E_{\sigma_0} + {\cal A}_{\sigma} (t)\right ) Z_{\sigma} (t, s) \Psi 
\eeq

\noi where $H'_{\sigma}$ is given by (\ref{2.17e}).}\\

\noi {\bf Proof.} We first remark that all the operators in the product
defining $Z_{\sigma}(t,s)$ leave ${\cal D}(H_0)$ invariant by
Proposition 2.2, part 2, by Lemma 4.1, part 2 and by Proposition 2.1.
From the fact that $C(\chi_{\sigma}\alpha_{\lambda} (t))^*$ is strongly
differentiable in ${\cal D}(H_0)$ by Lemma 4.1, part 3 and that
$U_{\sigma} (t)$ is strongly differentiable in ${\cal D}(H_0)$ by
Proposition 2.1, it follows that $Z_{\sigma} (t,s) \Psi$ is
differentiable and that its time derivative is given by 
\bea
\label{4.16e}
i\ {d \over dt} Z_{\sigma}(t,s) \Psi &=& Q_{\sigma}^* \Big \{ - i a^* \left ( \chi_{\sigma}  \dot{\alpha}_{\lambda}(t)\right )  + i a \left ( \chi_{\sigma}  \overline{\dot{\alpha}}_{\lambda}(t)\right ) + \ {\rm Im} \left ( \chi_{\sigma}  \alpha_{\lambda}(t), \dot{\alpha}_{\lambda}(t)\right ) \nn \\
&& + C  \left ( \chi_{\sigma}  \alpha_{\lambda}(t)\right )^* \left ( H_{\sigma} - p E_{\sigma}\right ) C \left ( \chi_{\sigma}  \alpha_{\lambda}(t)\right )\Big \} Q_{\sigma} \ Z_{\sigma} (t,s) \Psi \ .\nn \\
\eea 

\noi Using (\ref{4.4e}), (\ref{4.5e}) and (\ref{4.9e}), we continue (\ref{4.16e}) as 
$$\cdots = Q_{\sigma}^* \Big \{ - i a^* \left ( \chi_{\sigma}  \dot{\alpha}_{\lambda}(t)\right )  + i a \left ( \chi_{\sigma}  \overline{\dot{\alpha}}_{\lambda}(t)\right ) + \ {\rm Im} \left ( \chi_{\sigma}  \alpha_{\lambda}(t), \dot{\alpha}_{\lambda}(t)\right )$$
$$+ H_{\sigma} - p E_{\sigma} + a^* \left ( \chi_{\sigma}\omega \alpha_{\lambda}(t)\right ) + a  \left ( \chi_{\sigma} \omega \overline{\alpha_{\lambda}(t)}\right ) $$
$$+ \ \parallel \chi_{\sigma} \omega ^{1/2}\alpha_{\lambda}(t)\parallel^2 \ + {\cal A}_{\sigma}(t) \Big \} Q_{\sigma}\  Z_{\sigma} (t,s) \Psi$$
$$= Q_{\sigma}^* \Big \{ H_{\sigma} - p E_{\sigma}+ {\cal A}_{\sigma}(t) \Big \} Q_{\sigma} \ Z_{\sigma} (t,s) \Psi$$

\noi which yields (\ref{4.15e}) by (\ref{2.15e}) and the fact that
$Q_{\sigma}$ commutes with ${\cal A}_{\sigma}(t)$. \par \nobreak \hfill
$\sq$ \par

The operator $H'_{\sigma} - p E_{\sigma_0} + {\cal A}_{\sigma} (t)$ contains
only positive powers of $\lambda$ and, as a form, is equivalent to $(1
+ H_0)$ uniformly in $\lambda$ for $\lambda$ sufficiently small. More
precisely we have the following lemma. \\

\noi {\bf Lemma 4.2.} {\it Let $\alpha \in L_{1/2}^2$ and let $\alpha
(t)$ be given by (\ref{4.3e}). Let $\lambda_0$ and $\sigma_0$ be as in
Proposition 2.4. Then there exist two constants $\rho$ and $C$ such that
\beq
\label{4.17e}
C^{-1}\parallel (1 + H_0)^{1/2}\Psi \parallel^2\ \leq \ <\Psi, \left ( \rho + H'_{\sigma} - p E_{\sigma_0} + {\cal A}_{\sigma} (t) \right ) \Psi > \ \leq \ C \parallel (1 + H_0)^{1/2}\Psi \parallel^2 
\eeq

\noi for all $\sigma$ with $\sigma_0 < \sigma$, $\lambda$ with
$|\lambda | < \lambda_0$, $t \in {I\hskip-1truemm R}$ and $\Psi \in {\cal D}(H_0^{1/2})$.
The constants $\rho$ and $C$ depend on $\alpha$ through the norm
$\parallel \omega^{1/2} \alpha \parallel_2$.}\\

\noi {\bf Proof.} The estimate (\ref{3.6e}) implies
$$\left | < \Psi, {\cal A}_{\sigma}(t) \Psi > \right | \ \leq \varepsilon <\Psi, H_0\Psi > \ + \ C \varepsilon^{-1/3} \parallel \omega^{1/2} \alpha \parallel_2^{4/3} \ \parallel \Psi \parallel^2$$

\noi which together with Proposition 2.4, part 1 and the definition
(\ref{2.16e}) of $E_{\sigma_0}$ yields (\ref{4.17e}). \par \nobreak
\hfill $\sq$ \par
 
Using Lemma 4.2 we now prove that $Z_{\sigma}(t,s)$ satisfies a uniform
boundedness property and has a strong limit when $\sigma$ tends to
infinity.\\

\noi { \bf Proposition 4.2} {\it Let $\alpha \in L_{1/2}^2$ and let
$\alpha (t)$ be given by (\ref{4.3e}). Let $\lambda_0$ and $\sigma_0$
be as in Proposition 2.4. Then\par

1) For any compact interval $I$ there exists a constant $C_I$ such that
\beq
\label{4.18e}
\parallel (1 + H_0)^{1/2} Z_{\sigma}(t,s) \Psi \parallel \ \leq \ C_I \parallel (1 + H_0)^{1/2}\Psi \parallel
\eeq

\noi for all $\sigma$ with $\sigma_0 < \sigma$, $\lambda$ with $|\lambda
| \leq \lambda_0$, $t, s \in I$ and $\Psi \in {\cal D}(H_0^{1/2})$.
The constant $C_I$ depends on $\alpha$ through the norm $\parallel
\omega^{1/2} \alpha \parallel_2$. \par

2) For any $t, s$ the following strong limit exists
\beq
\label{4.19e}
{\rm s} - \lim_{\sigma \to \infty} Z_{\sigma} (t, s) = Q^* W(t,s) Q \equiv Z(t,s)
\eeq

\noi and $Z(t,s)$ satisfies the same estimate (\ref{4.18e}) as $Z_{\sigma}(t,s)$.}\\

\noi {\bf Proof.} \underbar{Part 1} We know already by Proposition 2.2,
part 2, by Lemma 4.1, part 2 and by Proposition 2.1 that
$Z_{\sigma}(t,s) {\cal D} (H_0^{1/2}) = {\cal D}(H_0^{1/2})$. Let 
$$M_{\sigma} (t) \equiv \rho + H'_{\sigma} - p E_{\sigma_0} + {\cal A}_{\sigma}(t)$$

\noi where $\rho$ is the constant that appears in Lemma 4.2 and let
$\Psi \in {\cal D}(H_0)$. The function $<Z_{\sigma}(t,s) \Psi,
M_{\sigma}(t) Z_{\sigma}(t,s) \Psi >$ is differentiable in the variable
$t$. In fact the differentiability of $Z_{\sigma}(t,s)$ is known by
Proposition 4.1 and the differentiability of $M_{\sigma}(t)$ is a
consequence of the fact that $\partial_t M_{\sigma}(t) = \dot{\cal
A}_{\sigma}(t)$. In addition since $A_{\sigma}$ and $\dot{A}_{\sigma}$
belong to $L^{\infty} ({I\hskip-1truemm R}^3)$, the operators ${\cal
A}_{\sigma} (t)$ and $\dot{\cal A}_{\sigma}(t)$ are bounded in ${\cal H}$ and
strongly continuous in $t$. Therefore
\beq
\label{4.20e}
{d \over dt}\ <Z_{\sigma} (t,s) \Psi, M_{\sigma}(t)Z_{\sigma} (t,s) \Psi > \ = \ <Z_{\sigma} (t,s)\Psi, \dot{\cal A}_{\sigma}(t) Z_{\sigma} (t,s)\Psi > 
\eeq

\noi and by integration
\bea
\label{4.21e}
&&<Z_{\sigma} (t,s) \Psi , M_{\sigma}(t)Z_{\sigma} (t,s) \Psi > \ = \ <\Psi, M_{\sigma}(s) \Psi >\nn \\
&&+ \int_s^t dt' \ <Z_{\sigma} (t',s) \Psi, \dot{\cal A}_{\sigma}(t')Z_{\sigma} (t',s) \Psi> \ .
\eea

\noi Using the estimate (\ref{4.17e}) for the terms with $M_{\sigma}$
and the estimate (\ref{3.8e}) for the term with $\dot{\cal A}_{\sigma}$
we obtain
$$\parallel (1 + H_0)^{1/2}Z_{\sigma}(t,s) \Psi \parallel^2\ \leq \ C \Big ( \parallel (1 + H_0)^{1/2}\Psi \parallel^2\ $$
$$+ \left | \int_s^t dt' <Z_{\sigma} (t',s) \Psi, H_0^{3/4} Z_{\sigma} (t',s) \Psi> \right | \Big )$$

\noi which yields trivially the linear inequality
\bea
\label{4.22e}
&&\parallel (1 + H_0)^{1/2}Z_{\sigma}(t,s) \Psi \parallel^2\ \leq \ C \Big ( \parallel (1 + H_0)^{1/2}\Psi \parallel^2\nn\\
&&+ \left | \int_s^t dt' \parallel (1 + H_0)^{1/2}Z_{\sigma} (t',s) \Psi \parallel^2 \right | \Big ) \ .
\eea

\noi By integrating (\ref{4.22e}) we obtain (\ref{4.18e}) for $\Psi
\in {\cal D}(H_0)$. We then extend (\ref{4.18e}) to all $\Psi \in {\cal
D} (H_0^{1/2})$ by continuity.\\

\noi \underbar{Part 2}. By Proposition 2.2, part 3, by Lemma 4.1, part
1 and by Proposition 2.4, part 3, all operators in the product
(\ref{4.13e}) of $Z_{\sigma}(t,s)$ converge strongly when $\sigma$
tends to infinity. The estimate (\ref{4.18e}) for $Z(t,s)$ follows from
that convergence and from the uniformity of the estimate for
$Z_{\sigma}(t,s)$ in $\sigma$. \par \nobreak \hfill $\sq$\par

We are now in condition to take the limit $\lambda \to 0$. \\

\noi {\bf Proposition 4.3.} {\it  Let $\alpha \in L_{1/2}^2$, let
$\alpha (t)$ be given by (\ref{4.3e}) and let $A$ be defined by
(\ref{3.3e}). Let $W(t,s)$ be defined by (\ref{4.2e}) and let $V(t,s)$
be defined in Proposition 3.1. Then the following strong limit exists
\beq
\label{4.23e}
{\rm s}  - \lim_{\lambda \to 0} W(t,s) = V(t,s)
\eeq

\noi uniformly for $t,s$ in compact intervals.} \\

\noi {\bf Proof.} Let $I$ be a compact interval and let $t,s \in I$.
Let $\lambda_0$ and $\sigma_0$ be as in Proposition 2.4. Let $\sigma >
\sigma_0$ and $\Psi \in {\cal D}(H_0)$. We estimate the difference
$$\parallel \left ( Z_{\sigma}(t,s) - V(t,s)\right )\Psi \parallel^2 \ = 2{\rm Re} \left \{ <\Psi, \Psi> - <Z_{\sigma}(t,s)\Psi,  V(t,s)\Psi > \right \}$$
$$= - 2 {\rm Re} \int_s^t dt' {d \over dt'}\ < Z_{\sigma}(t',s)\Psi,  V(t',s) \Psi > =$$
$$= 2 {\rm Im} \int_s^t dt' \Big \{ <\left ( H'_{\sigma} - p E_{\sigma_0} + {\cal A}_{\sigma}(t')\right ) Z_{\sigma}(t',s) \Psi , V(t',s)\Psi >$$
$$- <Z_{\sigma}(t',s)\Psi,  \left ( H_0 + {\cal A}(t')\right ) V(t',s)\Psi > \Big \}$$
\beq
\label{4.24e}
= 2{\rm Im} \int_s^t dt' <Z_{\sigma}(t',s)\Psi, \left ( ( H'_{\sigma}  - H_0)- p E_{\sigma_0} + {\cal A}_{\sigma} (t') - {\cal A}(t')\right ) V(t',s)\Psi > 
\eeq

\noi where we have used Proposition 3.1 and Proposition 4.1. We now
apply the estimate (\ref{3.12e}) to $V(t,s)$ and the estimate
(\ref{4.18e}) to $Z_{\sigma}(t,s)$, thereby obtaining
$$\parallel \left ( Z_{\sigma}(t,s) - V(t,s)\right )\Psi \parallel^2 \ \leq C_I \int_s^t dt' \Big \{ \parallel (1 + H_0)^{-1/2} (H'_{\sigma} - H_0) (1 + H_0)^{-1/2}\parallel$$
\beq
\label{4.25e}
+ p |E_{\sigma_0}| + \ \parallel (1 + H_0)^{-1/2} \left ( {\cal A}(t') - {\cal A}_{\sigma} (t')\right ) (1 + H_0)^{-1/2} \parallel \Big \}  \parallel(1 + H_0)^{1/2} \Psi \parallel^2
\eeq

\noi with $C_I$ uniform in $\sigma > \sigma_0$ and in $\lambda$, $|\lambda | \leq \lambda_0$. Now
$$\parallel (1 + H_0)^{-1/2} (H'_{\sigma} - H_0) (1 + H_0)^{-1/2}\parallel \ \leq C \lambda \ ,$$
$$ p |E_{\sigma_0}| \leq C \lambda^2$$

\noi by Lemma 2.2 and by (\ref{2.16e}) respectively, and 
$$\parallel (1 + H_0)^{-1/2} \left ( {\cal A}_{\sigma}(t') - {\cal A}(t')\right ) (1 + H_0)^{-1/2} \parallel \ \leq \ C \parallel (1 + H_0)^{-1} H_0^{1/4} \parallel \ \parallel (1 - \chi_{\sigma}) \omega^{1/2} \alpha \parallel_2$$

\noi by the estimate (\ref{3.6e}), so that (\ref{4.25e}) implies 
\beq
\label{4.26e}
\parallel \left ( Z_{\sigma}(t,s) - V(t,s)\right )\Psi \parallel^2 \ \leq C_I \left ( \lambda + \ \parallel (1 - \chi_{\sigma})\omega^{1/2} \alpha \parallel_2 \right ) \parallel ( 1 + H_0)^{1/2} \Psi \parallel^2
\eeq

\noi with $C_I$ uniform in $\lambda$, $|\lambda | \leq \lambda_0$ and
in $\sigma > \sigma_0$. Taking the limit $\sigma \to \infty$ in
(\ref{4.26e}) and using Proposition 4.2, part 2, we obtain 
\beq
\label{4.27e}
\parallel \left ( Z(t,s) - V(t,s)\right )\Psi \parallel \ \leq C_I \ \lambda^{1/2} \parallel ( 1 + H_0)^{1/2} \Psi \parallel \ .
\eeq

\noi From the identity
$$W(t,s) - V(t,s) = Q\ Z(t,s) Q^* - V(t,s) $$
$$= (Q-1) Z(t,s) + Q \ Z(t,s) (Q^*-1) + Z(t,s) - V(t,s)$$

\noi we obtain the estimate
\bea
\label{4.28e}
 \parallel \left ( W(t,s) - V(t,s) \right ) \Psi \parallel & \leq & \parallel (Q-1) Z(t,s) \Psi \parallel \ + \ \parallel (Q^* - 1)\Psi \parallel\nn \\
&&+\ \parallel \left ( Z(t,s) - V(t,s) \right ) \Psi \parallel \ .
\eea

\noi Now
$$\parallel (Q^* - \1)\Psi \parallel \ \leq \ \parallel T \Psi \parallel\ \leq \ \parallel T(1 + H_0)^{-1/2}\parallel \ \parallel (1 + H_0)^{1/2} \Psi \parallel$$

\noi so that from the definition (\ref{2.13e}) of $T$ and from (\ref{2.6e}) (\ref{2.7e}), 
\beq
\label{4.29e}
\parallel (Q^* - \1)\Psi \parallel \ \leq \ C \ \lambda \parallel (1 + H_0)^{1/2} \Psi \parallel
\eeq

\noi where the linear dependence in $\lambda$ comes from the linear dependence of $T$ on $g$ and therefore on $\lambda$.\par

Similarly from (\ref{4.29e}) using (\ref{4.18e}) for $Z(t,s)$ we estimate 
\beq
\label{4.30e}
\parallel (Q - 1)Z(t,s)\Psi \parallel \ \leq \ C_I \ \lambda \parallel (1 + H_0)^{1/2} \Psi \parallel\ .
\eeq

\noi By substituting (\ref{4.27e}), (\ref{4.29e}) and (\ref{4.30e}) into (\ref{4.28e}) we obtain 
\beq
\label{4.31e}
\parallel (W(t,s) - V(t,s))\Psi \parallel \ \leq \ C_I \ \lambda^{1/2} \parallel (1 + H_0)^{1/2} \Psi \parallel\ .
\eeq

\noi This proves the convergence of $W(t,s) \Psi$ to $V(t,s) \Psi$ for
any $\Psi \in {\cal D}(H_0)$ when $\lambda$ converges to zero,
uniformly for $t,s$ in compact intervals. Convergence for any $\Psi \in
{\cal H}$ follows from the unitarity of $W(t,s)$ and $V(t,s)$. \par
\nobreak \hfill $\sq$ \par

\end{document}